\documentclass[aps,prb,twocolumn,groupedaddress,showpacs]{revtex4}
\usepackage{graphicx}
\usepackage{dcolumn}
\usepackage{bm}

\begin{document}

\title{Angular-dependent oscillations of the magnetoresistance in 
Bi$_{2}$Se$_{3}$ due to the three-dimensional bulk Fermi surface}

\author{Kazuma Eto}
\author{Zhi Ren}
\author{A. A. Taskin}
\author{Kouji Segawa}
\author{Yoichi Ando}
\affiliation{Institute of Scientific and Industrial Research, 
Osaka University, Ibaraki, Osaka 567-0047 Japan}


\begin{abstract}

We observed pronounced angular-dependent magnetoresistance (MR)
oscillations in a high-quality Bi$_2$Se$_3$ single crystal with the
carrier density of 5$\times$10$^{18}$ cm$^{-3}$, which is a topological
insulator with residual bulk carriers. We show that the observed
angular-dependent oscillations can be well simulated by using the
parameters obtained from the Shubnikov-de Haas oscillations, which
clarifies that the oscillations are solely due to the bulk Fermi
surface. By completely elucidating the bulk oscillations, this result
paves the way for distinguishing the two-dimensional surface state in
angular-dependent MR studies in Bi$_2$Se$_3$ with much lower carrier
density. Besides, the present result provides a compelling demonstration
of how the Landau quantization of an anisotropic three-dimensional Fermi
surface can give rise to pronounced angular-dependent MR oscillations. 

\end{abstract}




\maketitle 

\section{Introduction}

The three-dimensional (3D) topological insulator is a rapidly growing
field of research in the condensed matter physics.
\cite{K1,MB,K2,SCZ1,Majorana,Monopole,Exiton,TYN,K4,SCZ3,SCZ4,H1,H2,H3,
H4,H5,Shen,Matsuda,AliY,Xue,QO,Ong,YiCui,SM,Furu,Vish,DHL,Fisher,AMRO}
Because the bulk of a topological insulator belongs to a different $Z_2$
topological class \cite{K1,MB,K2} than the vacuum, on the surface of a
3D topological insulator emerges an intrinsically metallic
two-dimensional (2D) state which hosts helically spin-polarized Dirac
fermions. \cite{K1,K2} Besides the profound implications of those
spin-helical 2D Dirac fermions on future spintronics, the topological
insulator is expected to present various exotic quantum phenomena
associated with its non-trivial topology.
\cite{K2,SCZ1,Majorana,Monopole,Exiton,TYN}

Motivated by theoretical predictions, \cite{K2,SCZ4} three materials
have so far been experimentally confirmed
\cite{H1,H2,H3,H4,H5,Shen,Matsuda} to be 3D topological insulators:
Bi$_{1-x}$Sb$_x$, Bi$_2$Se$_3$, and Bi$_2$Te$_3$. The angle-resolved
photoemission experiments \cite{H1,H2,H3,H4,H5,Shen,Matsuda} have played
decisive roles in indentifying the topologically nontrivial nature of
their surface states. More recently, scanning tunneling spectroscopy
experiments \cite{AliY,Xue} have elucidated the protection of the
surface state from spin-nonconserving scattering. In addition to those
surface-sensitive probes, transport experiments are obviously important
for understanding the macroscopic properties of topological insulators
and for exploiting the applications of their novel surface state.
However, capturing a signature of the surface state of a topological
insulator in its transport properties has proved difficult: The surface
state has been seen by quantum oscillations only in Bi$_{1-x}$Sb$_x$ at
a particular Sb concentration of 9\%; \cite{QO} the universal
conductance fluctuations associated with the surface state were observed
only in Ca-doped Bi$_2$Se$_3$ after careful tuning of the carrier
density; \cite{Ong} the Aharonov-Bohm oscillations through the surface
state were observed only in very narrow nanoribbons of Bi$_2$Se$_3$.
\cite{YiCui} Most often, the transport properties are dominated by the
bulk conductivity due to residual carriers.\cite{Ong,Fisher}

In this context, we have very recently found \cite{AMRO} that the 2D
surface state of the topological insulator Bi$_{0.91}$Sb$_{0.09}$ gives
rise to new types of oscillatory phenomena in the angular-dependence of
the magnetoresistance (MR). It was demonstrated \cite{AMRO} that the
oscillations observed at lower fields provide a way to distinguish the
2D Fermi surface (FS) in Bi$_{0.91}$Sb$_{0.09}$, while another type of
oscillations of unknown origin was found to become prominent at higher
fields. In the present work, we have tried to elucidate whether this new
tool, the angular-dependence of the MR, can distinguish the 2D surface
state in Bi$_2$Se$_3$, which is believed to be the most promising
topological insulator for investigating the novel topological effects
because of its simple surface-state structure. \cite{SCZ4,H3} We
observed pronounced angular-dependent oscillations of the MR in an
$n$-type Bi$_2$Se$_3$ single crystal with the carrier density $n_e$ =
5$\times$10$^{18}$ cm$^{-3}$, but our detailed analysis clarified that
the observed oscillations are solely due to the Landau quantization of
the anisotropic bulk FS of this material. Nevertheless, to the best of
our knowledge, such an angular-dependent MR oscillations in a 3D
material has never been explicitly demonstrated in the literature, so
the present results nicely supplement our general understanding of the
angular-dependent MR oscillation phenomena. In addition, our analysis
provides a solid ground for discriminating the contributions of the 2D
and 3D FSs in the angular-dependence of the MR in future studies of
Bi$_2$Se$_3$ single crystals with much lower carrier density.

\section{Experimental Details}

High-quality Bi$_{2}$Se$_{3}$ single crystals were grown by melting
stoichiometric mixtures of 99.9999\% purity Bi and 99.999\% purity Se
elements in sealed evacuated quartz tubes. After slow cooling from the
melting point down to about 550$^{\circ}$C over two days, crystals were
kept at this temperature for several days and then were furnace-cooled
to room temperature. The obtained crystals are easily cleaved and reveal
a flat shiny surface. The X-ray diffraction measurements confirmed the
rhombohedral crystal structure of Bi$_{2}$Se$_{3}$. Post-growth
annealings at various temperatures under controlled selenium partial
pressures were used to reduce the number of selenium vacancies that are
responsible for creating electron carriers. \cite{Ong,Fisher} 

The resistivity $\rho_{xx}$ was measured by a standard four-probe method
on rectangular samples, with the electric current $I$ directed along the
$C_1$ axis. Continuous rotations of the sample in constant magnetic
field $B$ were used to measure the angular dependence of the transverse
MR within the $C_3$-$C_2$ plane. For some selected magnetic-field
directions, the field dependences of $\rho_{xx}$ and the Hall
resistivity $\rho_{yx}$ were also measured by sweeping $B$ between
$\pm$14 T.

\begin{figure}
\includegraphics*[width=18pc]{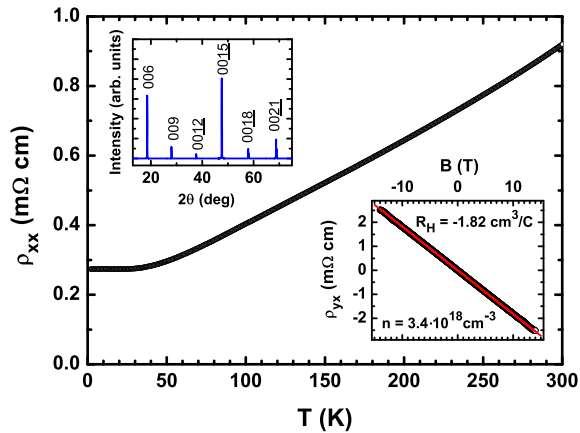}
\caption{(Color online) 
Temperature dependence of $\rho_{xx}$ in 0 T. 
Upper inset shows the X-ray diffraction pattern of the Bi$_2$Se$_3$ 
single crystal used for transport measurements.
Lower inset shows $\rho_{yx}$ for $B \parallel C_3$ measured at 1.5 K. 
The slope of $\rho_{yx}(B)$, shown by the thin solid line, 
suggests that the main carriers are electrons whose density is 
3.4$\times$10$^{18}$ cm$^{-3}$.
}
\label{fig1}
\end{figure}

\section{Results and Discussions}

\subsection{Resistivity and SdH Oscillations}

Figure 1 shows the temperature dependence of $\rho_{xx}$ of the
Bi$_{2}$Se$_{3}$ single crystal studied in this work. It shows a
metallic behavior $d\rho/dT > 0$ down to $\sim$30 K, and saturate at
lower temperature (there is actually a weak minimum near 30 K, as is
usually observed \cite{Fisher,Kohler0} in low-carrier-density
Bi$_{2}$Se$_{3}$). The single-crystal nature of the sample is evident
from the X-ray diffraction data shown in the upper inset of Fig. 1. The
lower inset of Fig. 1 shows the Hall resistivity $\rho_{yx}$ measured at
1.5 K for the field direction along the $C_3$ axis, which suggests that
the main carries are electrons and the carrier density $n_e$ is
3.4$\times$10$^{18}$ cm$^{-3}$ (in a one band model). From the values of
the Hall coefficient $R_H$ = 1.82 cm$^3$/C and $\rho_{xx}$ = 0.28
m$\Omega$cm at 1.5 K, the Hall mobility $\mu_H$ is estimated to be 6500
cm$^2$/Vs.

\begin{figure}
\includegraphics*[width=17pc]{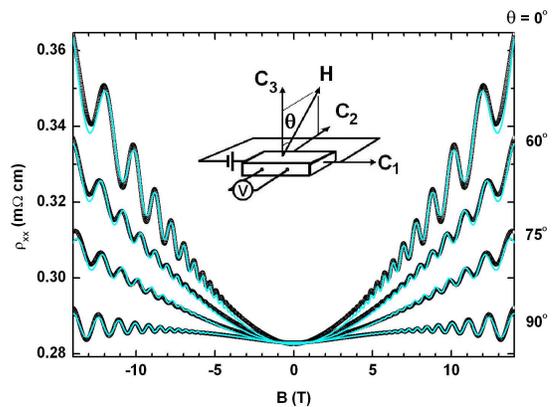}
\caption{(Color online) 
SdH oscillations measured within the $C_3-C_2$ plane. 
Thin solid lines are the result of our $\rho_{xx}(B)$ simulation 
(see text). Inset shows the measurement configuration.
}
\label{fig2}
\end{figure}

Figure 2 shows $\rho_{xx}(B)$ measured at 1.5 K for several field
directions in the transverse geometry ($I \parallel C_1$ and $B \perp
I$), after removing the antisymmetric components due to the leakage of
$\rho_{yx}$. Two features are evident: First, pronounced Shubnikov-de
Haas (SdH) oscillations are seen for any field direction, suggesting
their 3D origin. Second, the background of the SdH oscillations varies
significantly with the field direction, indicating that the transverse
MR is very anisotropic. Both features are taken into account in our
simulation of the observed angular dependence of the MR, as will be
discussed below.

\begin{figure}
\includegraphics*[width=20.5pc]{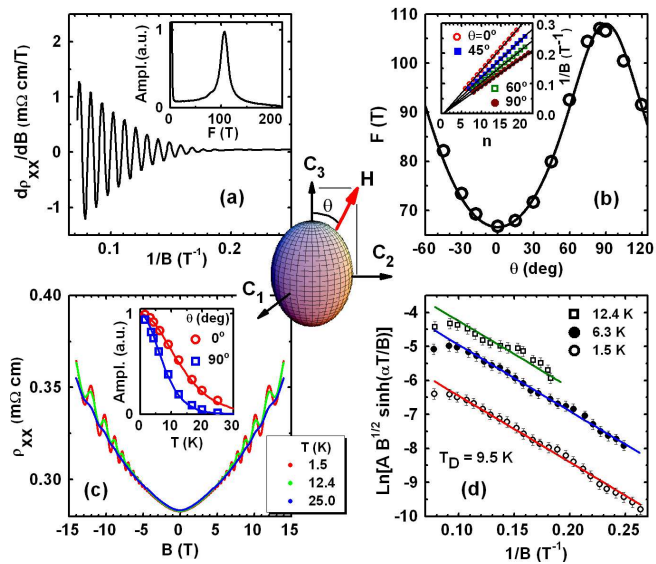}
\caption{(Color online) 
Analyses of the SdH oscillations. 
The inset in the middle schematically shows the obtained 3D Fermi surface
and the definition of $\theta$.  
(a) SdH oscillations for $B \parallel C_2$ as a function of $1/B$. 
The Fourier transform shown in the inset reveals a single frequency 
$F$ = 107 T.
(b) $F(\theta)$ measured within the $C_3-C_2$ plane; inset shows the 
fan diagram for several $\theta$.
(c) Temperature dependence of the SdH oscillations for $B \parallel C_3$; 
inset shows the temperature dependences of the SdH amplitudes measured
along the $C_3$ and $C_2$ axes at 12 T, yielding the cyclotron mass
of 0.14$m_e$ and 0.24$m_e$, respectively.
(d) Dingle plots for $B \parallel C_3$ at several temperatures give the 
same $T_D$ = 9.5 K.
}
\label{fig3}
\end{figure}

Figure 3 presents the analysis of the observed SdH oscillations. The
oscillations in $d\rho_{xx}/dB$ plotted as a function of $1/B$ for $B
\parallel C_2$ ($\theta = 90^{\circ}$) are shown in Fig. 3(a) as an
example. The very simple pattern seen in Fig. 3(a) is a result of the
single frequency $F$ = 107 T (see inset for the Fourier transform)
governing the SdH oscillations.\cite{note1} The same analysis was
applied to the data for other field directions, and the obtained $F$ as
a function of $\theta$ are shown in Fig. 3(b). The same set of
frequencies can be extracted from the Landau-level ``fan diagram" [inset
of Fig. 3(b)], which is a plot of the positions of maxima in
$\rho_{xx}(B)$ as a function of the Landau level numbers. The slopes of
the straight lines in the fan diagram give exactly the same $F(\theta)$
as the Fourier transform result. Another piece of information that can
be extracted from the fan diagram is the phase of the oscillations,
$\gamma$, which is determined by $\rho_{xx} \sim \cos[2 \pi (F/B
+\gamma)]$. In the present case, all the straight lines in the inset of
Fig. 3(b) intersect the horizontal axis at the same point, giving
$\gamma$ = 0.4 that is independent of the field direction. The angular
dependence of the SdH frequency $F(\theta)$ points to a single
ellipsoidal FS located at the $\Gamma$ point with the semi-axes $k_{a}$
= $k_{b}$ = 4.5$\times$10$^{6}$ cm$^{-1}$ ($\perp C_3$) and $k_{c}$ =
7.3$\times$10$^{6}$ cm$^{-1}$ ($\parallel C_3$). The expected
$F(\theta)$ for this FS is shown by the solid line in Fig. 3(b), which
fits the data very well. The carrier density corresponding to this FS is
5$\times$10$^{18}$ cm$^{-3}$, which is about 50\% higher than that
obtained from $R_H$, as was reported previously in the literature.
\cite{Kohler1} This discrepancy is due, at least partly, to the fact
that the present $R_H$ was measured in the low-field limit and contains
the so-called Hall factor, which is usually between 1 -- 2.

The temperature dependence of the SdH oscillations was measured for two
field directions, along the $C_3$ and $C_2$ axes. Figure 3(c) shows the
$\rho_{xx}(B)$ data in $B \parallel C_3$ for some selected temperatures,
where one can see that the background MR is essentially
temperature-independent and that oscillations are still visible even at
25 K. The inset of Fig. 3 (c) shows the temperature dependence of the
SdH amplitude measured at 12 T for $B$ along $C_3$ ($\theta$ =
0$^{\circ}$) and $C_2$ ($\theta$ = 90$^{\circ}$). The fits with the
standard Lifshitz-Kosevich theory \cite{Shnbrg} yield the cyclotron mass
$m_c$ of 0.14$m_e$ and 0.24$m_e$ for $\theta$ = 0$^{\circ}$ and
90$^{\circ}$, respectively. The energy dispersion near the conduction
band minimum of Bi$_2$Se$_3$ is known to be parabolic, \cite{Kohler1}
and the observed anisotropy in $m_c$ for the given FS is entirely
consistent with the parabolic dispersion to within 5\%. Relying on this
dispersion, the Fermi energy $E_F$ is calculated to be 56 meV and the
Fermi velocity $v_F$ is 3.8$\times$10$^{7}$ cm/s for any direction. 

The Dingle plots [shown in Fig. 3(d) for $B \parallel C_3$ as an
example] yield the Dingle temperature $T_D$ = 9.5 K which is almost
isotropic and is constant at low temperature; this gives an isotropic
scattering time $\tau$ = 1.3$\times$10$^{-13}$ s, implying the mean free
path $\ell$ (= $v_F \tau$) of $\sim$50 nm and the mobility $\mu_{\rm
SdH}$ of about 1600 cm$^{2}$/Vs, which is almost four times smaller than
$\mu_H$ estimated from $R_H$. Such a discrepancy between $\mu_{\rm SdH}$
and $\mu_H$ has long been noted in various systems, \cite{Shnbrg} and
the essential reason lies in the difference in how the effective
scattering time comes about from the microscopic scattering process:
\cite{DasSarma} The scattering rate $1/\tau_{\rm tr}$ determining
transport properties acquires the additional factor $1-\cos\phi$ upon
spatial averaging ($\phi$ is the scattering angle), while $1/\tau$ to
govern the dephasing in the quantum oscillations is given by a simple
spatial averaging without such a factor. Hence, if the small-angle
scattering becomes dominant (which is often the case at low
temperature), $1/\tau_{\rm tr}$ can be much smaller than $1/\tau$. This
means that the mean free path in our sample can be even larger than that
estimated above.

\begin{figure}
\includegraphics*[width=17pc]{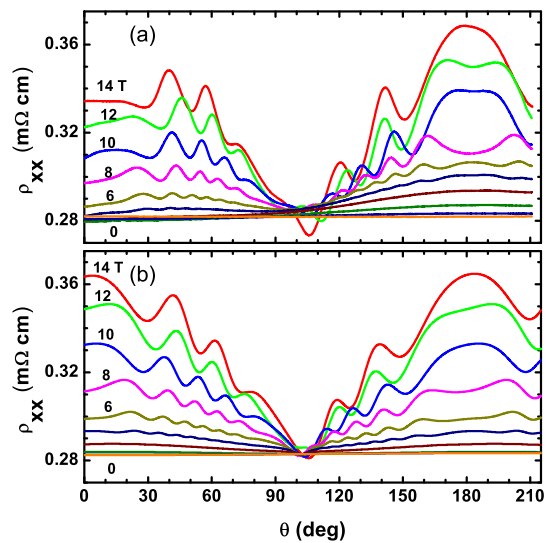}
\caption{(Color online) 
(a) Angular dependences of $\rho_{xx}$ measured 
within the $C_3-C_2$ plane in constant magnetic fields, whose values
were 0, 0.5, 1.5, 3, 4.5, 6, 8, 10, 12, and 14 T.
(b) Simulation of $\rho_{xx}$ based on Eq. (1) for the same magnetic 
fields as in (a).
}
\label{fig4}
\end{figure}

\subsection{Angular-Dependent MR Oscillations}

The new observation of the present work is the oscillatory
angular-dependence of the MR shown in Fig. 4(a), where, on top of the
two-fold-symmetric background angular dependence, pronounced
oscillations are evident at higher fields. Importantly, the peak
positions shift with magnetic field, which is different from the
ordinary angular-dependent MR oscillations (AMRO) in
quasi-low-dimensional systems. \cite{Yakovenko} In the following, we
show that these oscillations are due to the Landau quantization of the
3D FS and, hence, are essentially of the same origin as the SdH
oscillations. The SdH oscillations occur as the Landau-quantized
cylinders in the Brillouin zone expand and cross the FS with increasing
magnetic field, while the angular-dependent oscillations occur as the
axis of those cylinders rotates in the Brillouin zone. Obviously, no
oscillation is expected for rotating magnetic field when the FS is
spherical, but when the FS is anisotropic, the number of cylinders
residing within the FS changes as the cylinder axis is rotated, leading
to resistivity oscillations. Therefore, an anisotropy in the FS is a
necessary ingredient for the angular-dependent oscillations.

\begin{figure}
\includegraphics*[width=17pc]{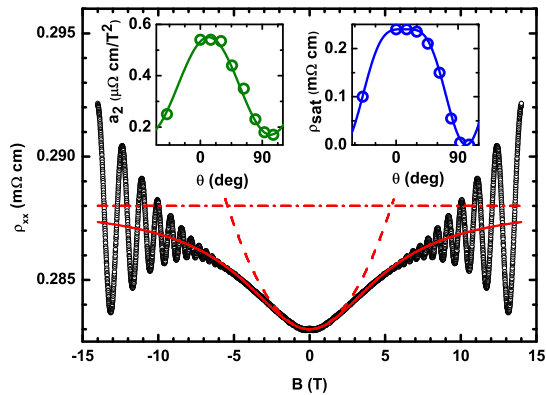}
\caption{(Color online) 
Fitting of the background MR for $B \parallel C_2$. Open circles show
the experimental data [symmetrical component of $\rho_{xx}(B)$]; the
dashed line demonstrates the low-field quadratic dependence of the
resistivity, $\rho_{\rm qd}(B) = a_2B^2$; the dash-dotted line
represents the high-field saturation limit, $\rho_{\rm sat}$; the solid
line is the resulting reconstructed background, $\rho_{\rm BG}(B)$ (see
text). The left and right insets show the obtained angular dependences
of the coefficients $a_2$ and $\rho_{\rm sat}$ within the $C_3-C_2$
plane.
}
\label{fig5}
\end{figure}

To show that the observed angular-dependent oscillations are essentially
due to the Landau quantization, it is most convincing to reconstruct the
angular-dependence of the MR based on the SdH oscillations data. For
this purpose, one needs to understand the exact magnetic-field
dependence of the MR in the presence of the SdH effect, and know how it
evolves when the magnetic field is rotated. Figure 5 shows the
$\rho_{xx}(B)$ data for $B \parallel C_2$ ($\theta = 90^{\circ}$), which
we take as an example for presenting our procedure to extract the
necessary information. At low magnetic fields, the MR in Fig. 5 exhibits
an almost quadratic field dependence, which can be fitted by $\rho_{\rm
qd}(B) = a_2B^2$. With increasing magnetic field, the MR tends to
saturate, while pronounced SdH oscillations develop at the same time.
The overall background MR $\rho_{\rm BG}(B)$ can be described as
$1/\rho_{\rm BG} = 1/\rho_{\rm qd} + 1/\rho_{\rm sat}$, which combines
the low-field quadratic behavior and the high-field saturation of
$\rho_{xx}(B)$. This background develops on top of the zero-field
resistivity $\rho_{xx}(0)$ of about 0.28 m$\Omega$cm. The same fitting
of the background of MR can be made for the whole range of the
magnetic-field direction. The evolutions of the parameters $a_2$ and
$\rho_{\rm sat}$ with $\theta$ in the $C_3-C_2$ plane are shown in the
left and right insets of Fig. 5. Note that the center of the symmetry
for both $a_2(\theta)$ and $\rho_{\rm sat}(\theta)$ is shifted from
$\theta$ = 0$^{\circ}$ by about 12$^{\circ}$, which is probably due to
the rhombohedral symmetry of the crystal which leads to the appearance
of cross terms \cite{Zitter} in the magnetic-field expansion of the
resistivity tensor.\cite{note2}

Once $\rho_{\rm BG}$ is known as functions of both $\theta$ and $B$, one
can superpose the SdH oscillations to obtain the expected
$\rho_{xx}(\theta, B)$. In the case of our Bi$_2$Se$_3$, thanks to the
fact that the SdH oscillations are composed of only one frequency, one
can simulate experimental data as
\begin{eqnarray}
\rho_{xx}(\theta, B) &=& \rho_{\rm BG}(\theta, B) \times \nonumber \\
                     & & \left(1+A R{_T} R{_D} R{_S} \cos \left[ 2\pi 
\left(\frac{F(\theta)}{B}+\gamma \right) \right] \right) , \nonumber \\
                     & &
\end{eqnarray} 
where $R_T$, $R_D$ and $R_S$ are temperature, Dingle, and spin damping
factors, respectively. \cite{Shnbrg} As shown in Fig. 2 by thin solid
lines, the experimentally observed SdH oscillations are reproduced
reasonably well with Eq. (1) with a parameter $A \cdot R_S$ = 0.12 which
is independent of $\theta$. 

The same set of parameters can be used to calculate the
angular-dependent MR for fixed $B$, $\rho_{xx}(\theta)$, with Eq. (1).
The result is shown in Fig. 4(b), where one can easily see that the
calculated angular dependences of the MR follow very closely the rather
complicated patterns of the observed $\rho_{xx}(\theta)$. This gives
compelling evidence that the angular-dependent oscillations are
essentially due to the Landau quantization of the anisotropic 3D FS.

It was emphasized in Ref. \onlinecite{AMRO} that a merit of measuring
the angular-dependence of the MR is that at sufficiently high magnetic
field, the 3D FS remains in the quantum limit (where all the electrons
condense into the 1st Landau level) and do not contribute to the
angular-dependent oscillations, while the 2D FS will always produce MR
oscillations when the magnetic-field direction is nearly parallel to the
2D plane. Once the density of the residual carriers in a topological
insulator sample is sufficiently reduced, the 3D FS should be easily
brought into the quantum limit and the 2D FS would become
distinguishable in the angular-dependent MR oscillations. Also, the SdH
effect needs a certain range of magnetic field for the oscillations to
be recognized, whereas the observation of the angular-dependent
oscillations can be made only at the highest field; this gives a certain
practical advantage to the latter method when the quantization condition
is only barely achieved with the available magnetic field. Therefore,
one would expect that the angular-dependence of the MR will be a useful
tool for distinguishing the 2D surface state of the topological
insulator Bi$_2$Se$_3$ when a single crystal with much lower carrier
density becomes available.

\section{Conclusions}

In conclusion, we observed pronounced angular-dependent oscillations of
the MR in high-quality single crystals of $n$-type Bi$_2$Se$_3$ with
$n_e$ = 5$\times$10$^{18}$ cm$^{-3}$. We show, by simulating the
angular-dependent oscillations based on the information obtained from
the SdH analysis, that the oscillations are essentially due to the
Landau quantization of the 3D bulk Fermi surface. This provides a
compelling demonstration of how the Landau quantization of an
anisotropic 3D FS can give rise to pronounced angular-dependent MR
oscillations. Furthermore, the present results pave the way for
distinguishing the 2D surface state in Bi$_2$Se$_3$ in future studies of
the angular-dependent MR, by completely elucidating the oscillations due
to the bulk FS.

\begin{acknowledgments}
This work was supported by JSPS (KAKENHI 19340078 and 2003004) and AFOSR 
(AOARD-08-4099).
\end{acknowledgments}


\begin{thebibliography}{99}
\bibitem{note1}
Strictly speaking, there is a subtle shoulder structure near 80 T in the
Fourier transform shown in the inset of Fig. 3(a), which suggests that
there may be a weak contribution from a second frequency. In fact, it is
known \cite{Kohler1} for Bi$_2$Se$_3$ that the second conduction-band
minima begin to be populated when $n_e$ becomes high, and in our sample
$n_e$ is actually close to this onset. Nevertheless, the possible
contribution of this second electron pocket in our sample is so weak
that it has no consequence in the transport properties being analyzed.
\end{thebibliography}
\end{document}